\documentclass[prl,twocolumn,superscriptaddress,showpacs,showkeys]{revtex4}
\usepackage{graphics}
\usepackage{color}
\usepackage{dcolumn}
\usepackage{pslatex}
\usepackage{amsmath}
\usepackage{amssymb}

\include{MyMc} 

\begin{document}

\title{%
  High efficiency deterministic Josephson Vortex Ratchet
}

\author{M.~Beck}
\author{E.~Goldobin}
\email{gold@uni-tuebingen.de}
\affiliation{%
  Physikalisches Institut -- Experimentalphysik II,
  Universit\"at T\"ubingen,
  Auf der Morgenstelle 14,
  72076 T\"ubingen, Germany
}
\author{M.~Neuhaus}
\author{M.~Siegel}
\affiliation{
 Universit\"at Karlsruhe,
 Institut f\"ur Mikro-- und Nanoelektronische Systeme,
 Hertzstr. 16,
 D-76187 Karlsruhe, Germany
}
\author{R.~Kleiner}
\author{D.~Koelle}
\affiliation{%
  Physikalisches Institut -- Experimentalphysik II,
  Universit\"at T\"ubingen,
  Auf der Morgenstelle 14,
  72076 T\"ubingen, Germany
}




\date{\today}

\begin{abstract}

We investigate experimentally a Josephson vortex ratchet --- a fluxon
in an asymmetric periodic potential driven by a deterministic
force with zero time average. The highly asymmetric periodic potential is created in an underdamped annular long Josephson junction by means of a current injector providing efficiency of the device up to 91\%. We measured the ratchet effect for driving forces with different spectral content. For monochromatic high-frequency drive the rectified voltage becomes quantized. At high driving frequencies we also observe chaos, sub-harmonic dynamics and voltage reversal due to the inertial mass of a fluxon. 

\end{abstract}

\maketitle



\Q{short par below}
The ratchet effect, \ie, the net unidirectional motion of a particle in a spatially asymmetric periodic potential in the presence of deterministic or stochastic forces with zero time average, received a lot of attention during the 20-th century. The second law of thermodynamics does not allow to extract useful work out of equilibrium thermal fluctuations, as was didactically demonstrated by Feynman\cite{Feynman:Lectures:Par46}. 
Thus, the only way to produce useful work is to supply non-native fluctuations (usually colored noise), which is the basic principle of operation for any ratchet.


Particularly during the last decade ratchets were receiving a lot of attention\cite{
Juelicher:1997:MMM,Reimann:2002:BrownianMotors,Haenggi:2005:BrownMotors}. Several new implementations, in particular based on the motion of the Josephson phase in SQUIDs\cite{Sterck:SQUID-Ratchets} or vortices in long Josephson junctions (LJJ)\cite{Goldobin:RatchetT:2001,Carapella:RatchetT:2001,Carapella:RatchetE:2001,Carapella:2002:JVR-HighFreq} or Josephson junction arrays (JJA)\cite{Falo:1999:JJA-Ratchet,Trias:2000:JJA-Ratchet,Lee:2003:JJA-Ratchet}, were suggested and tested. The investigation of quantum ratchets%
\cite{Linke99a,Grifoni:2002:QuRatWithFewBands,Majer:QuJVR},
\ie, a quantum particle moving/tunneling quantum mechanically in an asymmetric potential, is a fascinating new field not very well developed up to now especially experimentally. Advantages of Josephson junction based ratchets are: (I) directed motion results in an average dc voltage which is easily detected experimentally; (II) Josephson junctions are very fast devices which can operate (capture and rectify noise) in a broad frequency range from dc to $\sim 100 \units{GHz}$, thus capturing a lot of spectral energy; (III) by varying junction design and bath temperature both overdamped and underdamped regimes are accessible; and (IV) one can operate Josephson ratchets in the quantum regime\cite{Majer:QuJVR}.

In this letter we investigate experimentally the \emph{deterministic underdamped Josephson vortex ratchet} (JVR), in which a Josephson vortex (fluxon) moves along a LJJ. We implemented a novel, effective way to construct a strongly asymmetric potential by means of a current injector and systematically study a quasi-statically driven ratchet with different spectral content of the driver. For non-adiabatic drive we observe quantized rectification, \emph{voltage reversal}, sub-harmonic, and chaotic dynamics.


Our system can be described by the following perturbed sine-Gordon equation\cite{Goldobin:RatchetT:2001}
\begin{equation}
  \phi_{xx}-\phi_{tt}-\sin\phi = \alpha\phi_t - \gamma(x) -\xi(t)
  , \label{Eq:sG}
\end{equation}
where $\phi$ is the Josephson phase, the curvilinear coordinate $x$ along the LJJ and the time $t$ are normalized to the Josephson penetration depth $\lambda_J$ and inverse plasma frequency $\omega_p^{-1}$, accordingly, $\alpha$ is the dimensionless damping parameter and $\gamma(x)=j_\mathrm{inj}(x)/j_c$ and $\xi(t)=j(t)/j_c$ are the bias current densities normalized to the critical current density of the LJJ. $\gamma(x)$ has zero spatial average and is used to create an asymmetric potential (see below). $\xi(t)$ is a spatially homogenous deterministic (or stochastic) drive with zero time average. The ultimate aim of ratchet operation is to rectify $\xi(t)$ to produce non-zero voltage $\langle \phi_t \rangle \ne0$. In the absence of the \rhs, the solitonic solution of Eq.~(\ref{Eq:sG}) is a Josephson vortex (sine-Gordon kink) $\phi(x)=4 \arctan\exp\left[(x-x_0(t))/\sqrt{1-u^2} \right]$, situated at $x_0(t)$ and moving with velocity $u=dx_0(t)/dt$. The \rhs of Eq.~(\ref{Eq:sG}) is usually considered as a perturbation.\cite{McLoughlinScott} It does not change drastically the vortex shape, but defines its dynamics, \eg, equilibrium velocity\cite{McLoughlinScott}. Such an approximation essentially treats the vortex as a rigid object, and its dynamics can be reduced to the dynamics of a relativistic underdamped point-like particle\cite{Carapella:RatchetT:2001} (cf. non-relativistic case \cite{Borromeo:2002:DetRat}). In this terms the ratchet should rectify $\xi(t)$ to produce a nonzero average velocity $\langle u \rangle \ne 0$.

To build a JVR, a fluxon should move in an asymmetric periodic potential $U(x_0)$. Opposite to soliton\cite{Salerno:2002:SolitonRat} or string\cite{Marchesoni:1996:ThermalRat1+1D} ratchets where the \emph{asymmetric} potential $V(\phi)$ is a primitive function of the current-phase relation (CPR), in our case, the CPR is sinusoidal, \ie, $V(\phi)=1-\cos(\phi)$ is a \emph{symmetric} function of $\phi$. Instead, we construct an asymmetric periodic potential $U(x_0)$, which is a function of the fluxon \emph{coordinate}. In contrast to other JVRs 
\cite{Falo:1999:JJA-Ratchet,Trias:2000:JJA-Ratchet,Lee:2003:JJA-Ratchet,Carapella:RatchetT:2001,Carapella:RatchetE:2001,Carapella:2002:JVR-HighFreq}, we construct $U(x_0)$ using a single current injector. As we proposed earlier (see the last paragraph before Sec.~III in Ref.~\onlinecite{Goldobin:RatchetT:2001})
the current injection with profile $\gamma(x)$ is equivalent to an applied nonuniform magnetic field $h(x)$ such that $h_x(x)=-\gamma(x)$. Then the potential $U(x_0) \approx -2\pi w h(x_0)$, where $w$ is the LJJ's width\cite{Goldobin:RatchetT:2001}. The periodicity of the potential is provided by using an annular LJJ (ALJJ) and $\gamma(x)$ with zero spatial average.

If we apply $\gamma(x)$ using a current injector of width $\Dw$ situated at $x=x_\mathrm{inj1}$ and extract the current \emph{from the same electrode} along the rest of the LJJ, i.e. $\gamma(x)=\gamma_1$ for $|x-x_\mathrm{inj1}|<\Dw/2$ and $\gamma(x)=-\gamma_2$ for all other $x$, then $U(x_0)$ looks like an asymmetric saw-tooth potential with the steep slope $\propto\gamma_1$ and the gentle slope $\propto\gamma_2$. Note that, zero spatial average of $\gamma(x)$ requires $j_c\gamma_1\Dw\,w =j_c\gamma_2(L-\Dw)w=I_\mathrm{inj1}$, where $L$ is the LJJ circumference. By changing $I_\mathrm{inj1}$ one modulates the amplitude of the potential. This allows operation as a flashing ratchet too. Below we focus on the rocking ratchet, \ie, when the potential $\propto I_\mathrm{inj1}$ is (almost) constant while $\xi(t)\ne0$.

Similar systems, but with magnetic field induced potential $U(x_0)$, were already studied for the case of quasi-static deterministic and stochastic drive\cite{Carapella:RatchetE:2001,Carapella:RatchetT:2001} and for deterministic high frequency drive\cite{Carapella:2002:JVR-HighFreq}. Ratchetlike systems based on an asymmetry of the driver (rather than the potential) were also investigated\cite{Marchesoni:1986:HarmonicMix,Flach:2002:BrokenSymm,Ustinov:2004:BiHarmDriverRatchet}.


\begin{figure}[!tb]
  \scalebox{0.5}{\includegraphics{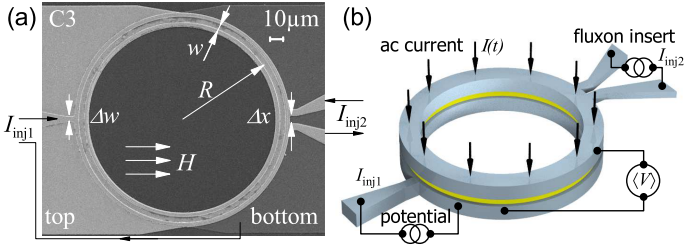}}
  \caption{
    (Color online) (a) Electron microscope image of sample C3.
    $R=70\units{\mu m}$, $\Delta w=5\units{\mu m}$, $\Delta x=5\units{\mu m}$. (b) Sketch of the ALJJ with inj1 and inj2. Wide top and bottom bias leads are not shown in (b) for clarity.
    }
  \label{Fig:sample1}
\end{figure}
%


%
\begin{table}[b]

 \begin{center}
  \begin{tabular}{|c|c|c|c|c|c|c|}
   \hline
   Sample & $R\units{[\mu m]}$ & $j_c\units{[A/cm^2]}$ & $\lambda_J\units{[\mu m]}$ 
   & $l$  & $\nu_0\units{[GHz]}$ & $V_1\units{[\mu V]}$ \\
   \hline
     G3& $130$ &  $97$ & $37$  & $21.8$ & $14$ & $28$  \\
     C2&  $70$ & $150$ & $30$  & $14.6$ & $24$ & $48$  \\
     C3&  $70$ &  $77$ & $42$  & $10.5$ & $24$ & $48$  \\
     C4&  $50$ & $165$ & $29$  & $10.9$ & $33$ & $66$  \\
    \hline
   \end{tabular}
  \end{center}
 \caption{Parameters of the used junctions. }
 \label{Tab:sample}
\end{table}

Experiments have been performed with Nb-AlO$_x$-Nb ALJJs that have geometry shown in Fig.~\ref{Fig:sample1}. A pair of current injectors (inj2) attached to the top electrode and separated by distance $\Dx$ can be used to insert a fluxon in the ALJJ.\cite{Ustinov:2002:ALJJ:InsFluxon,Malomed:2004:ALJJ:Ic(Iinj)}  A single injector (inj1) in the bottom layer is used to create the asymmetric potential as described above. We investigated several samples with different normalized circumferences $l=L/\lambda_J$ ($L=2\pi R$; $R$ is a mean radius, see Fig.~\ref{Fig:sample1}). The device parameters  are summarized in Tab.~\ref{Tab:sample}. 
For all samples $\lambda_J\gg w=5\units{\mu m}$ and $\lambda_J\gg \Dw=\Dx=5\units{\mu m}$ ($10\units{\mu m}$ for G3), \ie,  we can treat our ALJJ like a one dimensional LJJ and inj2 like an almost ideal discontinuity\cite{Gaber:Art2,Malomed:2004:ALJJ:Ic(Iinj)}. The quantity $\nu_0={\bar c}_{0}/L$ is the maximum revolution frequency of a fluxon (${\bar c}_{0}$ is the Swihart velocity) and $V_1=\Phi_0\nu_0=\Phi_0\omega_p/l$ is the corresponding voltage, \ie,  the asymptotic voltage of the first fluxon step.

%
Measurements were performed in a shielded cryostat at $T=4.2\units{K}$ unless stated otherwise. Before operating the ratchet, each ALJJ was characterized and injectors were calibrated.
All junctions listed in Tab.~\ref{Tab:sample} showed good $I$-$V$ characteristics (IVC) and nice symmetric $I_c(H)$ dependences (not shown). The dependence $I_c(I_\mathrm{inj2})$ looks like a Fraunhofer pattern (not shown) in accord with theory \cite{Malomed:2004:ALJJ:Ic(Iinj)}. The first minimum is reached at $I_\mathrm{inj2}=\pm3.9\units{mA}$ and corresponds to the phase twisted by $\pm2\pi$ in a tiny region between the injectors and to a free (anti)fluxon inserted into the ALJJ outside inj2.

\newcommand{\IcIinjIWithFluxon}{black symbols}
\newcommand{\IcIinjIOhneFluxon}{gray symbols}
\begin{figure}[tb]
  \includegraphics{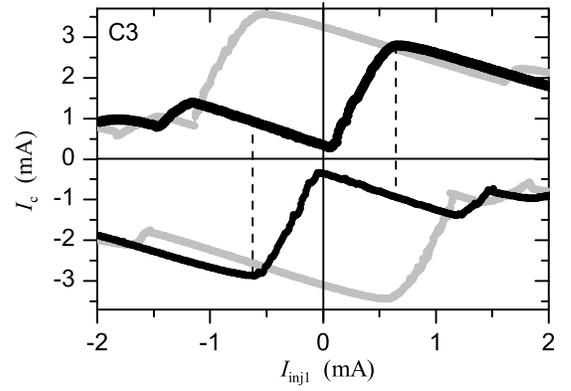}
  \caption{%
    The dependence $I_c(I_\mathrm{inj1})$ without trapped fluxon shows the asymmetry of equivalent field $h(x)$ (\IcIinjIOhneFluxon). The plot $I_c(I_\mathrm{inj1})$ with one trapped fluxon shows the asymmetry of potential $U(x_0)$ (\IcIinjIWithFluxon).
  }
  \label{Fig:IcIinj1}
\end{figure}

To calibrate inj1, we measure $I_c(I_{\rm inj1})$. The current $I_{\rm inj1}$ is applied between inj1 and the bottom electrode of the ALJJ. 
In Fig.~\ref{Fig:IcIinj1} {\IcIinjIOhneFluxon} show $I_c(I_{\rm inj1})$ with no fluxons trapped in the junction. Starting from the main maximum, the $I_c(I_{\rm inj1})$ curve has different slopes for increasing or decreasing $I_{\rm inj1}$. The ratio of these slopes corresponds to the ratio of slopes of the $h(x)$ saw-tooth\cite{Goldobin:RatchetT:2001}.  Then we applied $I_{\rm inj2}=\pm3.9\units{mA}$ to inject a fluxon, and measured $I_c(I_{\rm inj1})$ again, as shown by {\IcIinjIWithFluxon} in Fig.~\ref{Fig:IcIinj1}. We see that $I_c(0)$ dropped down 
almost to zero (the residual pinning is present due to the finite inj2 sizes $\Dx$ and $\Dw$\cite{Malomed:2004:ALJJ:Ic(Iinj)}). Applying a finite $I_{\rm inj1}$ creates a potential which pins the fluxon stronger and one needs to apply a larger bias current $I$ to let the fluxon move around the ALJJ and generate a voltage. For small $I_{\rm inj1}$ the depinning current $I_c$ grows almost linearly with $I_{\rm inj1}$, but it is \emph{asymmetric} for positive and negative direction of the bias current (driving force). This corresponds to different slopes of the potential when the fluxon tries to move to the left or to the right out of the well. The ratio of slopes of the $I_c(I_{\rm inj1})$ dependence for one trapped vortex reflects the asymmetry of the  potential $U(x_0)$\cite{Goldobin:RatchetT:2001} and for sample C3 is about $6.6$. Figure \ref{Fig:IcIinj1} also defines our ``working area'', \ie, the reasonable range of $|I_{\rm inj1}|\leq{0.6\units{mA}}$ for ratchet operation,  
shown by the dashed lines.


%
\begin{figure}[tb]
  \includegraphics{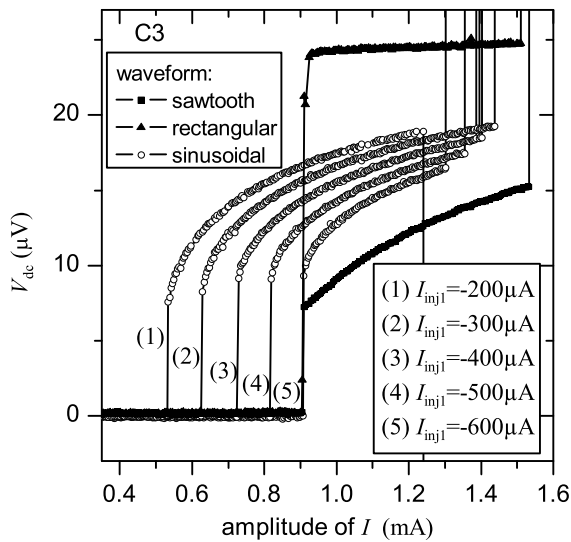}
  \caption{%
    Measured $V_{dc}(I_{ac})$ for different values of potential heights $\propto I_\mathrm{inj1}$ and for a sinusoidal (monochromatic) drive (open circles). Measured $V_{dc}(I_{ac})$ for $I_\mathrm{inj1}=-600\units{\mu A}$ and ac drives with different spectral content: rectangular  shape (solid triangles) and saw-tooth shape (solid squares).
  }
  \label{Fig:V_dc(I_ac)}
\end{figure}

If we apply $I_\mathrm{inj2}=-3.9\units{mA}$ (insert one fluxon) and $I_\mathrm{inj1}=-0.2\units{mA}$ (create a potential of intermediate amplitude) a fluxon step appears on the IVC (not shown), corresponding to the rotation of a fluxon around the ALJJ with $u\approx{\bar c}_{0}$. The depinning $I_c$ and return current $I_r$ of the fluxon step depends on the polarity of the applied bias current as well as on $I_{\rm inj1}$. To demonstrate rectification of the rocking ratchet we apply the periodic bias current (deterministic driving force) $I(t)=\xi(t)j_c L w=I_{ac}\sin(2\pi\nu t)$ with frequency $\nu=100\units{Hz}\ll\nu_0$ (quasi-static regime) and measure $V_{dc}(I_{ac})=\langle V \rangle (I_{ac})$ by averaging the voltage over $10\units{ms}$ ($1000$ data points sampled at $100\units{kHz}$) --- one period of the ac drive. For ALJJ C3 $V_{dc}(I_{ac})=V_1\cdot\langle u \rangle(I_{ac}) $ is shown in Fig.~\ref{Fig:V_dc(I_ac)} for different values of $I_\mathrm{inj1}$ (open circles). All $V_{dc}(I_{ac})$ curves have similar features. For small $I_{ac}$ the driving force acting on a fluxon is not sufficient to push the fluxon out of the well in either direction so that $V_{dc}=0$. At higher amplitude the bias is able to push the fluxon in one direction,  which results in $V_{dc}\ne 0$. The value of $I_{ac}$ at which rectification first appears grows linearly with $I_\mathrm{inj1}$ as it should be according to Fig.~\ref{Fig:IcIinj1}. 
Further, $V_{dc}$ grows with $I_{ac}$, but this dependence due to relativistic saturation of fluxon velocity is weaker than the linear dependence predicted for a non-relativistic particle\cite{Borromeo:2002:DetRat}.
At an even larger amplitude $I_{ac}$ the junction switches into the resistive state, generating a high positive or negative dc voltage. The latter regime is not discussed here as it has nothing to do with the operation of the JVR. We note that we very rarely observed a decrease of $V_{dc}$ at higher $I_{ac}$ --- a typical behavior for many ratchet systems. In our case, the asymmetry is so high, that a negative fluxon step does not appear in most cases. 

We have also investigated the influence of the shape (spectral content) of the driving force $I(t)$ on the performance of our ratchet. As an example, in Fig.~\ref{Fig:V_dc(I_ac)} we also show rectification curves $V_{dc}(I_{ac})$ for \emph{time symmetric} saw-tooth pulses and for a \emph{time symmetric} rectangular drive at $I_\mathrm{inj1}=-600\units{\mu A}$. One can see that rectangular pulses result in higher performance (rectification) because the ALJJ spends more time at the fluxon step than in the case of a sinusoidal drive. Similarly, a saw-tooth drive results in lower rectification. Usually it is believed that a drive with compact spectrum (\eg, monochromatic) is more efficient in terms of rectification than a drive with broad spectrum (\eg, white noise, which most frequently provides no rectification). Our results show that this is not always true: a rectangular drive with a rather broad discrete spectrum is more efficient than a monochromatic one. On the other hand, a saw-tooth drive also with broad spectrum, is less efficient than a sinusoidal one.

A typical value of rectified voltage $V_{dc}\approx 20\units{\mu V}\sim V_1/2$) and is higher than reported earlier\cite{Carapella:RatchetE:2001}. In principle, $V_1\propto 1/L$, so one can get higher $V_{dc}$ for a shorter LJJ. However, when the LJJ becomes too short, the asymmetry of the potential vanishes due to the convolution with the fluxon shape, see Eq.~(17) of Ref.~\onlinecite{Goldobin:RatchetT:2001}.

\begin{figure}[tb]
  \includegraphics{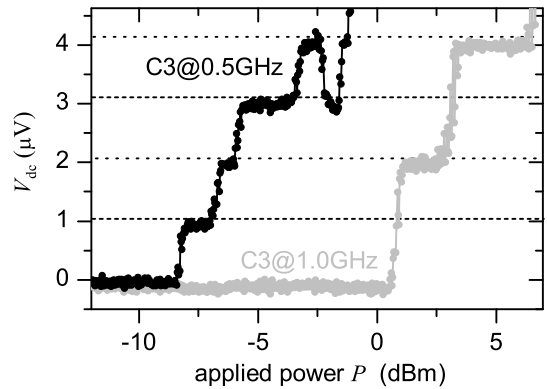}
  \caption{%
    Measured $V_{dc}(P)$ for $\nu=0.5\units{GHz}$ (black circles) and $\nu=1.0\units{GHz}$ (gray circles).
  }
  \label{Fig:RF1}
\end{figure}

\begin{figure}[tb]
  \includegraphics{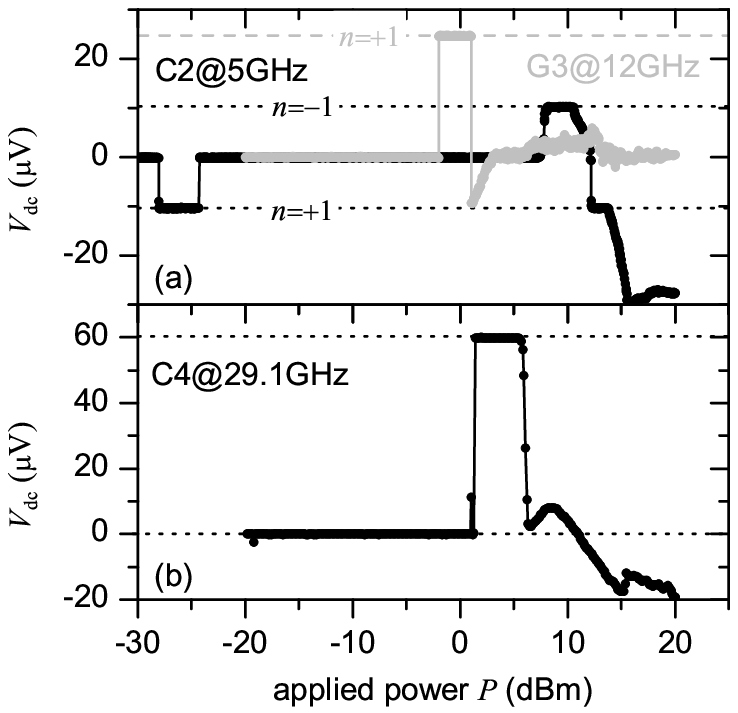}
  \caption{%
    Measured $V_{dc}(P)$. 
    (a) ALJJ C2 (black circles) $\nu=5\units{GHz}$, $I_\mathrm{inj1}=470\units{\mu A}$, $I_\mathrm{inj2}=5.3\units{mA}$ and
    ALJJ G3 (gray circles) $\nu=12\units{GHz}$, $I_\mathrm{inj1}=-400\units{\mu A}$, $I_\mathrm{inj2}=4.0\units{mA}$.
    (b) ALJJ C4 (black circles) $\nu=29.1\units{GHz}$, $I_\mathrm{inj1}=-580\units{\mu A}$, $I_\mathrm{inj2}=4.0\units{mA}$.
  }
 \label{Fig:RF2}
\end{figure}

To drive the ratchet at higher frequencies ($\sim 1\units{GHz}$) we placed an emitting antenna connected to an rf generator close to the ALJJ, so that the bias leads act as a pickup antenna. Note that this induces no signal in inj2 and a rather small signal in inj1, so that our ratchet has $90\ldots95\units{\%}$ of a ``rocking'' and $5\ldots10\units{\%}$ of a ``flashing'' potential. The dc voltage was averaged over $10\units{ms}$ ($\gtrsim 10^7$ periods) and recorded \vs applied power $P$ as shown in Fig.~\ref{Fig:RF1} for sample C3. Rather than growing smoothly, the dc voltage is quantized, $V_n=n\Phi_0\nu$ (Shapiro-like steps). Each step corresponds to an integer number $n$ of turns of a fluxon around the ALJJ per period of ac drive. The curves in Fig.~\ref{Fig:RF1} are quite noisy because their voltage is comparable to the residual noise of our measurement setup. Note that the $P$-axis shows the applied power at the output of the generator, so the curves corresponding to different frequencies may appear shifted along the $P$-axis due to the frequency dependent coupling of microwaves to the junction. At higher applied power we sometimes (\eg, for sample C3 at $\nu=0.5\units{GHz}$) see a stepwise decrease of the rectified dc voltage, see Fig.~\ref{Fig:RF1}. This feature can be reproduced in simulations and is related to the bifurcation to the period two dynamics. In general, multiple discrete voltage steps like in Fig.~\ref{Fig:RF1} can be observed for frequencies up to $3\units{GHz}$.

When the time $t_0$ required for one revolution of a fluxon around the ALJJ (in the best case $t_0={\bar c}_{0}/L$) becomes comparable with the period of the driving force $1/\nu$, \ie, the voltage $V=\Phi_0\nu$ is approaching $V_1$, the fluxon has time only for one revolution, so only one step can be observed on the $V_{dc}(P)$ curves. 

For frequencies $\nu>4\units{GHz}$ we used a $(0.7\times2.5\times1.5)\units{cm^3}$ copper box with the first mode at $6\units{GHz}$ to achieve better coupling and avoid multiple low frequency resonances. To make the system less chaotic we increase the damping by measuring at slightly higher temperature $T\approx6\units{K}$. In Fig.~\ref{Fig:RF2} we show $V_{dc}(P)$ for ALJJs C4, G3, and C2 (see Tab.~\ref{Tab:sample}). The curves show a single quantized voltage step corresponding to one fluxon revolution.
For sample C4 we achieved operation of the ratchet up to very
high frequency of $29\units{GHz}$ ($V_{dc}/V_1=\langle u \rangle =0.88$). For sample G3, which is longer, we reach $12\units{GHz}$ ($\langle u \rangle =0.91$). Our normalized average velocity $\langle u \rangle$ is considerably larger than 0.22\cite{Ustinov:2004:BiHarmDriverRatchet} or $0.33$\cite{Carapella:2002:JVR-HighFreq}, reported recently for similar ratchet(-like) systems. The curve for sample C2 shows a negative dc voltage step, as expected for
$I_\mathrm{inj1}=470\units{\mu A}>0$, see Fig.~\ref{Fig:IcIinj1}. Then $V_{dc}$ averages to zero over a large range of $P$, but for $7.5\units{dBm}<P<12\units{dBm}$ there appears a quantized dc voltage of opposite polarity. This \emph{voltage reversal} (or particle's \emph{current reversal}) corresponds to the motion of the fluxon in an unnatural direction and was predicted for point-like underdamped deterministically driven particle\cite{Barbi00,Jung1996:IntertiaRatchets,Mateos2000:UdampDetRat:CR}. Note the two ingredients essential for current reversal in our case: (a) high driving frequencies and (b) low damping. For low frequencies we observe no current reversal, see Fig.~\ref{Fig:V_dc(I_ac)}. In the overdamped regime a monochromatically driven ratchet shows no current reversal either\cite{Bartussek:1994:RockedThermRatchet}. At high power our JVRs show unquantized $V_{dc}$ regions, corresponding to chaotic dynamics. Such regimes can be reproduced in simulations. The details will be presented elsewhere.


In conclusion, we investigated experimentally \emph{a relativistic underdamped Josephson vortex ratchet} where a strongly asymmetric potential is created using a current injector. We observed quantized rectification of a deterministic signal at high frequencies up to $29\units{GHz}$, average velocity $\langle u \rangle$ up to $0.91$, voltage reversal, as well as sub-harmonic and chaotic regimes. Deterministic quasi-static signals with broad spectra may be rectified better or worse than monochromatic ones. 




We thank P. H\"anggi for discussions. This work is supported by the Deutsche Forschungsgemeinschaft.

\bibliographystyle{apsprl}
\bibliography{this,jj-annular,LJJ,SF,jj,software,ratch}

\end{document}